# Superconductivity of 30.4 K and its Reemergence under Pressure in Fe$_{1.11}$Se Synthesized via Ion-exchange and De-intercalation Reaction

Mingzhang Yang[†], Yuxin Ma[†], Qi Li, Ke Ma, Jiali Lu, Zhaolong Liu, Ruijin Sun, Tianping Ying, Mengdi Wang, Xin Chen, Changchun Zhao, Jian-gang Guo,* Shifeng Jin,* Xiaolong Chen*

*Email: jgguo@iphy.ac.cn; shifengjin@iphy.ac.cn; xlchen@iphy.ac.cn

**ABSTRACT:** Binary stoichiometry FeSe (s-FeSe) is a well-known parent of high-temperature unconventional superconductors owing to its charge-neutral layer, highly tunable structure and electronic properties, and rich interplays among multiple electronic phases. Yet the s-FeSe synthesized via high-temperature equilibrium reactions bears the notorious interstitial Fe, where merely 3% of them is sufficient to kill the superconductivity. Here, we successfully synthesized a new non-stoichiometric Fe$_{1.11}$Se single crystal with a superconducting onset temperature ($T_c^{onset}$) of 30.4 K through a hydrothermal ion-exchange and de-intercalation route. 11% interstitial Fe ions exceed the equilibrium phase diagram limit. Intriguingly, under physical pressure, the $T_c^{onset}$ of Fe$_{1+δ}$Se$_{1-x}$S$_x$ exhibits a "V"-shaped evolution with a minimum at 2–2.6 GPa, and then upturning into a second superconducting region, reminiscent of the behaviors in FeSe intercalates. Furthermore, a pressure-induced possible magnetic order, previously only observed in pressurized s-FeSe, shows up. These results offer fresh insights into the role of interstitial Fe in governing superconducting and transport properties under non-equilibrium synthesis and tuning strategies.

## INTRODUCTION

FeSe-based superconductors have become an exceptional member in the iron-based superconducting family on account of highly-tunable layered structures,[1–3] the highest critical temperature ($T_c$),[4] and topological surface state for potential applications in quantum computing,[5–7] *etc.* Stoichiometric FeSe (s-FeSe) synthesized by high-temperature sintering of 1000-1200 K, *i.e.* equilibrium-state synthesis method, exhibits a modest $T_c$ of 8.5 K.[8] Its superconductivity (SC) is extremely sensitive to Fe content, which is fully suppressed by mere 3 % interstitial Fe.[8] Partial substitution of Se by S/Te in s-FeSe slightly enhances $T_c$ to 12-15 K.[9–12] simultaneously bringing out competing or intertwining phenomena including magnetic fluctuations, quantum critical point, and rich electronic phases.[13–15]

In the past few years, to enhance $T_c$ and explore the superconducting mechanism of FeSe, researchers have dedicated major efforts to three categories of non-equilibrium approaches. The first involves chemical or electrochemical intercalation of structural units, by which alkali-metal ions (K⁺, Rb⁺, Cs⁺),[16–20] inorganic-organic molecule groups [Li$_x$(NH$_3$)$_y$, K$_x$(NH$_3$)$_y$, Na$_x$(NH$_3$)$_y$, Na$_x$(C$_2$H$_8$N$_2$)$_y$],[21–25] and inorganic hydroxide (Li$_y$Fe$_{1-y}$OH)$^{δ+}$ are inserted between FeSe layers.[26,27] These FeSe-intercalates exhibit the highest $T_c$ of 46 K, exceeding the McMillan limited temperature of the BCS theory framework. The second strategy is directly depositing one-unit-cell FeSe on SrTiO$_3$ substrate, in which much higher $T_c$ of 65-100 K are achieved.[4,28,29] The synergistic effect of interface charge transfer and phonon vibra-tion of oxygen is responsible for the remarkable enhancement of $T_c$. The third method is employing physical pressure to tune the atomic arrangements and configurations in favor of enhancing SC in a relatively simple way. The $T_c$ of s-FeSe increases monotonically to 37 K at 6-8 GPa, and then decreases, forming a dome-shaped superconducting zone.[30–33] Above 10 GPa, the superconducting anti-PbO FeSe changes into the non-superconducting NiAs phase. Notably, at 2-6 GPa, a new stripe-type antiferromagnetic (S-AFM) transition emerges, likely competing with the SC.[34,35] Besides, the FeSe-intercalates are also pressurized, and their $T_c$s can be enhanced above 50 K in re-entrant superconducting phases.[36–39] These studies indicate that effective electron doping and modifications to the FeSe$_4$ tetrahedral environment-either individually or jointly-play essential roles in enhancing SC and producing emergent electronic states.

Here, we develop a two-step hydrothermal ion-exchange and de-intercalation route to synthesize a new Fe$_{1.11}$Se bulk single crystals having 11% interstitial Fe2 ions, which is much higher than its solubility in the Fe-Se binary phase diagram. Remarkably, this new non-stoichiometric Fe$_{1.11}$Se exhibits a $T_c^{onset}$ of 30.4 K, identical to that in K$_x$Fe$_2$Se$_2$. The Fe2 thus acts as a benign dopant, which suppresses nematicity, injects electrons, and enhances SC. More interestingly, the $T_c$ is suppressed to a minimal value near 2.5 GPa in the SC-I zone, and then enhanced to a maximal value under pressure, forming a SC-II zone. In terms of superconducting evolution, such a V- shaped $T_c$ is totally different from single-dome SC in s-FeSe, while it resembles those of FeSe-intercalates. In the SC-II zone, an emergent possible magnetic phase coexists with SC above 5.6 GPa. The new pressure dependent electronic property seems bridge the gap between s-FeSe and FeSe-intercalates.

## RESULTS AND DISCUSSION

Single crystals of Fe$_{1+\delta}$Se$_{1-x}$S$_x$ ($x$=0, 0.3, 0.6) were successfully synthesized by the hydrothermal intercalation and selective de-intercalation procedure, see Figure 1a. Single-crystal X-ray diffraction (SCXRD) on a 152 × 66 × 11 μm$^3$ specimen at 299 K shows that Fe$_{1.11}$Se crystallizes in the LiFeAs-type structure with space group of $P4/nmm$.[40] Fe1 and Se fully occupy the 2b [1/4, 3/4, 0] and 2c [1/4, 1/4, 0.266(1)] Wyckoff sites. Differential Fourier maps reveal residual electron density between layers, attributed to interstitial Fe2 ions with an occupancy of 11 %, which randomly occupies the 2c site [1/4, 1/4, 0.866(1)]. More details are shown in Table S1 and Figure S1.

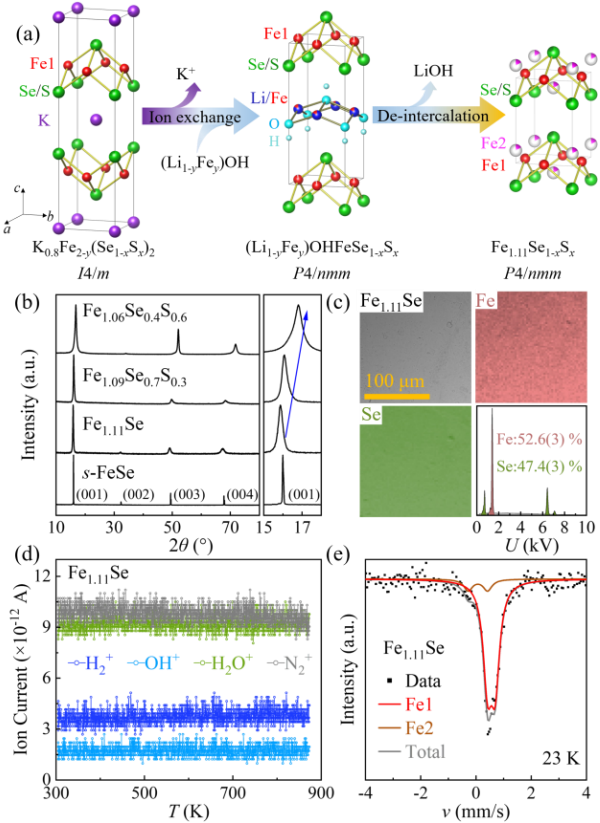

Figure 1. (a) Schematic synthesis route and crystal structure of Fe$_{1+\delta}$Se$_{1-x}$S$_x$. (b) Powder X-ray diffraction patterns of s-FeSe and Fe$_{1+\delta}$Se$_{1-x}$S$_x$ single crystals. (c) SEM image and EDX mapping of Fe$_{1.11}$Se. (d) Dynamic gas composition analysis of Fe$_{1.11}$Se. (e) $^{57}$Fe Mössbauer spectrum of Fe$_{1.11}$Se at 23 K.

The purity of single crystal form is also examined by Powder X-ray diffraction (PXRD) on a 0.5 × 0.3 cm$^2$ specimen, showing only sharp (00l) reflections and no detectable peaks of precursors K$_{0.8}$Fe$_{2-y}$(Se$_{1-x}$S$_x$)$_2$ and (Li$_{1-y}$Fe$_y$)OHFeSe$_{1-x}$S$_x$. Notably, the lowest-angle Bragg reflection corresponds to a $d$-spacing of 5.588(6) Å, 1.2% larger than that of s-FeSe. It implies the incorporation of additional species between adjacent FeSe layers. Upon sulfur doping, all (00l) peaks shift to higher angle. More details from refinement are listed in Table S2.

The elemental compositions of Fe$_{1.11}$Se crystals were determined using several characterization techniques. The complete de-intercalation of K$^+$ or Li$^+$ can be confirmed by inductively coupled plasma-atomic emission spectroscopy (ICP-AES) and energy-dispersive X-ray spectroscopy (EDX) (see Figure 1c). The ratio of Fe:Se measured by ICP-AES is 1.18(1):1. The EDX elemental mapping on freshly cleaved crystal surfaces (Figure 1c), which shows a homogeneous distribution of Fe and Se with the Fe:Se ratio of 1.11(1):1, in good agreement with the analysis of SCXRD measurements. The ratio of Fe:Se is determined as 1.09(1):1 and 1.06(1):1 for the $x$=0.3 and 0.6, respectively. We further performed *in-situ* dynamic gas composition analysis using a Quadrupole Mass Spectrometer (QMS). As shown in Figure 1d, the H$_2$$^+$ signal is 3.5 × 10$^{-12}$ A up to 870 K, which is smaller than the background N$_2$ signal. There are also negligible water-derived signals, i.e., OH$^+$ and H$_2$O$^+$, below 5 × 10$^{-12}$ A.

To further confirm the presence and local environment of Fe2 in Fe$_{1.11}$Se, we conducted a $^{57}$Fe Mössbauer spectrum at 23 K. The primary doublet (Fe1) exhibits an isomer shift (IS) of 0.56(1) mm·s$^{-1}$ and a quadrupole splitting of 0.28(1) mm·s$^{-1}$, which are nearly equal to that of s-FeSe (IS = 0.57(1) mm·s$^{-1}$, QS = 0.30(1) mm·s$^{-1}$ at 5 K).[8] The isomer shift of Fe2 ion in Fe$_{1.11}$Se is 0.07 mm·s$^{-1}$, which is lightly lower than the 0.14-0.2 mm·s$^{-1}$ of Fe2 ions in Fe$_{1.25}$Te. However, no additional doublets or magnetic splitting are observed in the spectrum, contrasting to the magnetic ordering of Fe$_{1+x}$Te ($x \geq 0.14$).[41,42] The interstitial Fe2 in Fe$_{1.11}$Se should be nomagnetic, and may locate at a less symmetric surrounding. Besides, the relative ratio of spectral areas for Fe1:Fe2 is 1:0.12(2), which quantitatively matches the 11% interstitial Fe2. Based on these observations, we can thus conclude that the existence of 11% Fe2 is reasonable. In addition, the effective moment of Fe2 is 4.9(1) $\mu_B$ ($S$=2, tetrahedral $d^6$) derived from temperature-dependent magnetic susceptibility as shown in Figure S2.

The electrical transport properties of Fe$_{1+\delta}$Se$_{1-x}$S$_x$ ($x$=0, 0.3, 0.6) were measured at ambient pressure as shown in Figure 2a. The $\rho$-$T$ curves of three samples do not have any kinks, suggesting the absence of nematic phase of s-FeSe. At lower temperatures, the sharp drop of $\rho$ due to superconducting transition emerges. We plot them in Figure 2b, where the $T_c$$^{onset}$ of Fe$_{1.11}$Se is 30.4 K, which is nearly four times higher than that of s-FeSe. While the $T_c$$^{onset}$ decreases to 17.4 K and 12.2 K in Fe$_{1.09}$Se$_{0.7}$S$_{0.3}$ and Fe$_{1.06}$Se$_{0.4}$S$_{0.6}$, respectively. We fitted the normal-state $\rho$-$T$ curves from the $T_c$$^{onset}$ to 80 K by $\rho(T) = \rho_0 + AT^\alpha$, as shown in Figure 1a. All $\alpha$ close to 1, indicating the non-Fermi-liquid behaviors. Magnetic measurements under ZFC conditions demonstrated the bulk SC in Fe$_{1.11}$Se. In Figure 2c, a sharp diamagnetic transition shows the $T_c$$^{zero}$ of 27.1 K at 10 Oe. The magnetization versus magnetic field (M-H) curves in Figure 2d exhibit a pronounced hysteresis loop at 5 K, characteristic of type-II superconductivity and strong flux-pinning. The $\rho$-$T$ curves below 50 K under different $H$ are presented in Figure S3. The "tail" of resistivity seems like the behavior in those 2D superconductors.[43,44] Figure 2e displays the linear field-dependent $\rho_{xy}$ up to 5 T at 50-200 K. It shows a single-band model dominated by electrons, which is different from the two-band model in *s*-FeSe. In all samples, the Hall coefficient $R_H$ are negative, as shown in Figure 2f. The electron concentration $n_e = 1/(R_H \cdot e)$ of $Fe_{1.11}Se$ is $1.28(1) \times 10^{21}$ cm$^{-3}$ at 50 K, which decreases upon doping sulfur. The $T_c$ also decreases as $n$ is lowered.

To probe the interlayer interaction in $Fe_{1.11}Se$, we measured the angular magnetoresistance of $Fe_{1.11}Se$ at 24 K up to 9 T (see Figure 2g). The SC is suppressed as the angle $\theta$ between the *c*-axis and the magnetic field increases from $-4°$ to $90°$, indicating anisotropic SC. Further analysis of the fluctuation-induced magnetoconductivity using the 2D Lawrence–Doniach (LD) model reveals that the coherence length ($\xi_c$) of $Fe_{1.11}Se$ along *c*-axis is 2.9 Å, remarkably shorter than $\xi_c$ of 26 Å in *s*-FeSe.[45-46] Thus $Fe_{1.11}Se$ is an anisotropic superconductor. Upon sulfur-doping, the $\xi_c$ increases to 5.5 Å and 18.1 Å for $x = 0.3$ and 0.6, and the anisotropic SC gradually weakens.

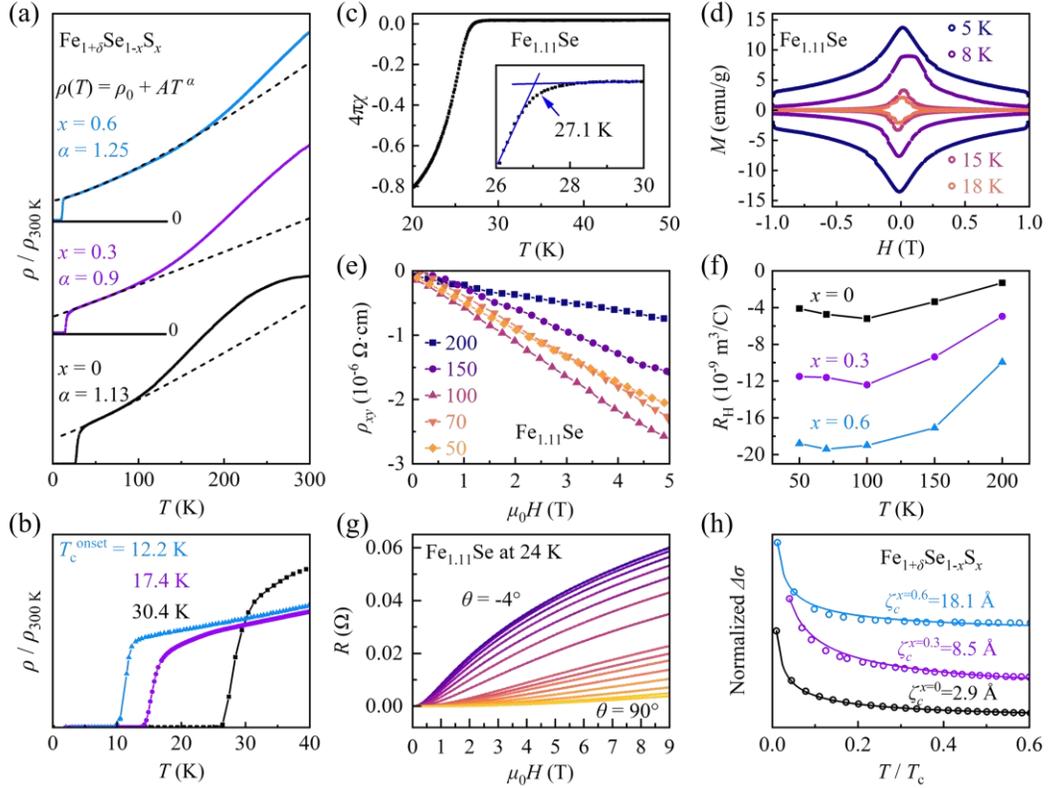

Figure 2. (a) Temperature dependence of normalized electrical resistivity ($\rho/\rho_{300\,K}$) of $Fe_{1+\delta}Se_{1-x}S_x$. Black, violet and blue curves are $x$=0, 0.3 and 0.6, respectively. Black dashed lines are the fitting curves. (b) Superconducting transition of $Fe_{1+\delta}Se_{1-x}S_x$. (c) Zero-field-cooled magnetization (ZFC) of $Fe_{1.11}Se$ at 10 Oe, and the inset is zoom-in transition. (d) *M-H* curves of $Fe_{1.11}Se$. (e) Hall resistivity of $Fe_{1.11}Se$ as a function of magnetic fields. (f) Temperature-dependent Hall coefficient $R_H$ of $Fe_{1+\delta}Se_{1-x}S_x$. (g) Magnetic field dependence of the resistance for $Fe_{1.11}Se$ at 24 K under varying angles. (h) Normalized $\Delta\sigma_{GT}$ derived from the Lawrence-Doniach (LD) model of three samples.

**Table 1. Comparison of $T_c$ and $n_e$ at 50 K in $Fe_{1.11}Se$, *s*-FeSe, and FeSe-intercalates.**

| | $Fe_{1.11}Se$ | *s*-FeSe[1,8,47] | $K_xFe_2Se_2$[16,48,49] | $(Li_{0.8}Fe_{0.2})OHFeSe$[26,38] | $Li_x(NH_3)_yFe_2Se_2$[21,43] |
|---|---|---|---|---|---|
| $T_c$ (K) | 30.4 | 8.5 | 30.1 | 40 | 44.3 |
| $n_e$ (×10$^{21}$ cm$^{-3}$) | 1.28(1) | 0.20(5) | 0.7-1.2 | 1.60(1) | 1.30(1) |

We freshly cleaved single crystals for *in-situ* electrical transport measurements under pressure. All the $\rho$-*T* curves are shown in Figures 3a and b. The $T_c^{onset}$ of $Fe_{1.11}Se$ initially decreases with applied pressure, reaching a minimum value of 15.4 K at 2.6 GPa in SC-I zone. Then it inversely increases to 31.1 K with further compression. As *P*>5.5 GPa, a weak kink of $\rho$-*T* curve emerges above $T_c^{onset}$. It may be a stripe-type antiferromagnetic (S-AFM) transition just observed in pressurized *s*-FeSe,[32,33,50,51] marked as violet arrow of $T_m$ in Figure 3b. This anomaly can be easily found in the temperature derivative d$\rho$/d*T* curves in Figure S4. Interestingly, both $T_m$ and $T_c^{onset}$ increase with increasing *P*, showing peaks of 35.2 K and 31.1 K at 10.6 GPa, respectively. At 12.5 GPa, the resistivity does not reach zero at 1.8 K. As *P*>13.6 GPa, neither superconducting nor magnetic transitions are observed. We also measured $\rho$-*T* curves of the sulfur-doped samples with $x = 0.3$ (Figures 3c and d) and 0.6 under pressure (see Figure S4). Their $T_c^{onset}$ show a similar V-shaped evolution with a minimal $T_c^{onset}$ at ~2.0 GPa. The $T_m$ transition also emerges as *P*>5.5 GPa. For $x$=0.3, the highest $T_c^{onset}$ is 25 K at 7.0 GPa. While in $x$=0.6, the highest $T_c^{onset}$ of reentered SC is 8 K at 5.5 GPa. In addition, the tetragonal-hexagonal phase transition in sulfur-doped samples decreases to 12 GPa and 10 GPa, which are reasonably due to the pre-exertion of chemical pressure by doping smaller sulfur ions in Fe$_{1.11}$Se.

To gain more insight into pressurized SC of Fe$_{1+\delta}$Se$_{1-x}$S$_x$, we plot the whole pressure-dependent electronic phase diagram in Figure 4a. In the three diagrams, the $T_c^{onset}$ and $T_c^{zero}$ decrease in the SC-I region and then lower to a minimum at $P$ of 2.0-2.6 GPa, followed by a dome-like $T_c$ in the SC-II. As $P$>13.6 GPa, 12 GPa, and 10 GPa, the SC of three samples totally disappears, respectively. We first examine whether the structural phase transition exists under pressure. The $in$-$situ$ synchrotron diffractions under pressure of $x$=0 and $x$=0.3 were conducted, and all the patterns are plotted in Figure S5. The contour plots clearly reveal that no structural transition occurs in two SC-I and SC-II zones. Further increasing pressure, additional peaks shup up, and the phases gradually change into a NiAs-type hexagonal structure. At the same time, the SC and possible magnetic phase transition disappear.

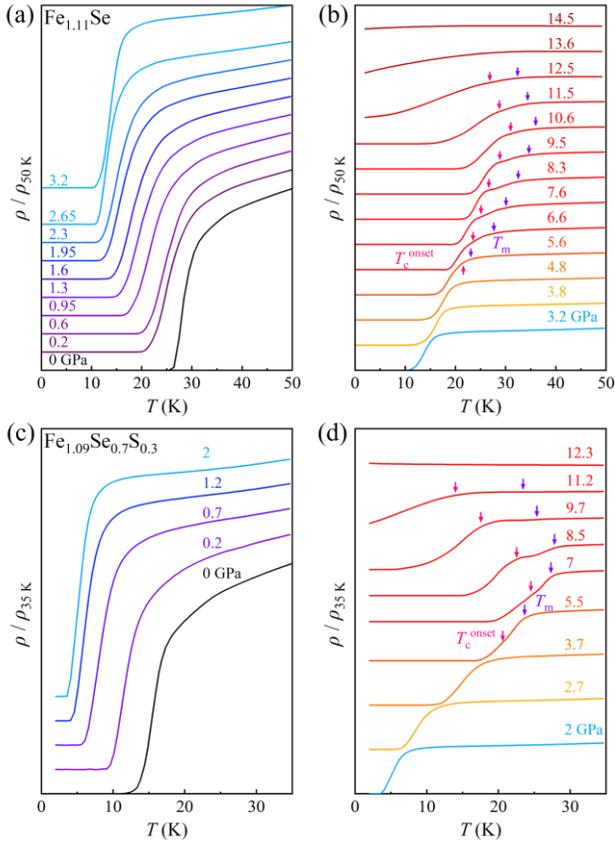

Figure 3. Temperature-dependent $\rho/\rho_{50\,K}$ and $\rho/\rho_{35\,K}$ in (a-b) Fe$_{1.11}$Se and (c-d) Fe$_{1.09}$Se$_{0.7}$S$_{0.3}$ single crystals, respectively, under different pressures. The red and violet arrows label the $T_c^{onset}$ and possible magnetic transition $T_m$, respectively.

In addition, we plot the pressure-dependent cell volume ratio $V/V_{0GPa}$ and lattice parameters ratio $c/a$ of Fe$_{1.11}$Se in Figures 4d and 4e, respectively. More details for Fe$_{1.09}$Se$_{0.7}$S$_{0.3}$ are plotted in Figure S6. For Fe$_{1.11}$Se, the $V/V_{0GPa}$ and $c/a$ ratio decrease smoothly and rapidly in 0-6 GPa, which are commonly observed in FeSe intercalates.[38,52] Above 6 GPa, the decrement of $V/V_{0GPa}$ seems to deviate from the standard Birch–Murnaghan (BM) equation of state, and the shrinkage of $c/a$ also slows down. These abnormal behaviors imply the formation of an interlayer bonding state under pressure.

Furthermore, we performed density functional theory (DFT) calculations using the lattice parameters of Fe$_{1.11}$Se under pressures. As shown in Figure 5by, the Fermi surface (FS) displays a quasi-2D cylindrical shape at 0 GPa. With increasing pressure, the bands along A-M cross the Fermi level, leading to the FS gradually warp toward a spherical form. At 10 GPa, the FS is fully enclosed within the first Brillouin zone, showing a 3D character. In the Figure S8a, we plotted the change of projected density of states and total density of states under pressure. It can be seen that the contributions of five orbitals almost do not change at $E_F$. As for the total value, it decreases from 21.42 states/eV to 17.54 states/eV as shown in Figure S8b, which do not directly relate to the value of $T_c$.

We listed the $T_c$ and $n_e$ of Fe$_{1.11}$Se, $s$-FeSe, and three FeSe intercalates in Table 1. The $n_e$ of Fe$_{1.11}$Se reached to a level comparable to those FeSe-intercalates with high $T_c$, indicating the doping effect of interstitial Fe. Furthermore, the Fe$_{1.11}$Se annealed at 200 °C still keeps the high $T_c^{onset}$ of 28.5 K, slightly lower than pristine Fe$_{1.11}$Se as plotted in Figure S9. However, the $T_c^{onset}$ of sample annealed at 400 °C moves down to 9.6 K identical to that of $s$-FeSe, indicating that Fe$_{1.11}$Se is a metastable superconductor sustained at low temperature.

Let's compare our $P$-$T$ phase diagrams with those in pressurized $s$-FeSe and FeSe-intercalates. In the former case, the nematic transition is suppressed by pressure, and the $T_c$ of 8 K slowly increases to 37 K at ~6 GPa. Detail $\rho$-$T$ measurements confirm there is a magnetic phase transition between 2- 6 GPa. As $P$>6 GPa, the $T_c$ slowly decreases and vanishes when a NiAs-type non-superconducting phase emerges. In K$_x$Fe$_2$Se$_2$, the $T_c$ of 31 K monotonically decreases to zero near 10 GPa, and then a second superconducting phase re-emerges with $T_c$ of 48 K.[36] This reentrance might be related to a quantum critical point.[37] In (Li$_y$Fe$_{1-y}$OH)FeSe and Li$_x$(NH$_3$)$_y$FeSe, a V-shaped SC is reported and the minimal $T_c$ occurs at 2-4 GPa.[38,39] However, the magnetic transition is not observed at all in these FeSe-intercalates. As for Fe$_{1.11}$Se, the pressure-dependent V-shaped SC is different from those of $s$-FeSe at lower pressure, while it is similar to the reports in FeSe-intercalates. In addition, at higher pressure region of 5.6-12.5 GPa, the emergent possible magnetic phase transition is alike the S-AFM transition in pressurized $s$-FeSe.

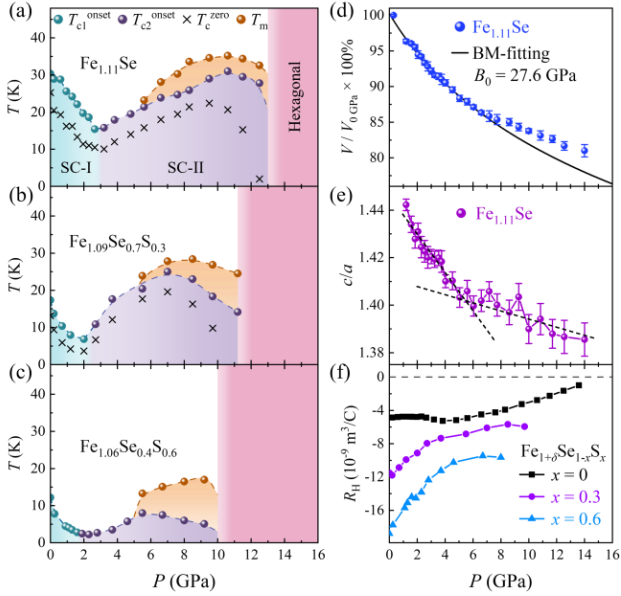

**Figure 4.** Phase diagram of pressurized $Fe_{1+\delta}Se_{1-x}S_x$ for (a) $x$=0, (b) $x$=0.3, and (c) $x$=0.6. (d) Derived cell volume ratio $V/V_{0\,GPa}$ and (e) lattice parameters ratio $c/a$ for $Fe_{1.11}Se$. The gray line is a fitted curve by the Birch-Murnaghan (BM) equation of state. The black dashed lines guide the trend of $c/a$. (f) Pressure dependence of $R_H$ at 50 K.

As previously reported, this magnetic transition is closely related to a hole-pocket located at the center of Brillouin zone of reconstructed Fermi surface. However, in $Fe_{1.11}Se$ and FeSe-intercalates, the doped electrons significantly enlarge the electron pockets. This is verified by the Hall measurements in $Fe_{1.11}Se$, where the $R_H$ of -4.86 × $10^{-9}$ $m^3$/C at ambient pressure equals the value of $(Li_yFe_{1-y}OH)FeSe$. Furthermore, all these values are negative, as shown in Fig. 4f, where the data were fitted by the single-band model (see Figure S10). Since the hole-pocket in $Fe_{1.11}Se$ should be pretty small, it is unlike the scenario of hole-pocket nesting induced magnetism in pressurized *s*-FeSe.[33] Alternately, one possible picture is the interstitial $Fe_2$ ions become magnetically interacted under pressure. Further studies will be of interest to clarify the origin.

In a recent work, the author claimed that the $T_c$ of FeSe based superconductors is positively correlated with electron concentrations.[47] This statement is supported by the relationship between $T_c$ and carrier concentrations of $Fe_{1+\delta}Se_{1-x}S_x$ at ambient pressure, as shown in Fig. 2f and Table 1. Nevertheless, the pressure-induced V-shaped SC is not fully explained by the pressure-dependent carrier concentrations, which monotonically increases with applied pressure. We did not observe critical points at 2- 2.6 GPa in Fig. 4f. Thus, additional parameters need to be taken into consideration.

In addition, double-dome SC was found in many iron based superconductors like the case of $LaFeAsO_{1-x}H_x$ and pressurized $K_xFe_2Se_2$.[36,37,53,54] In the former one, the two-dome structure is closely related to bipartite magnetic parents LaFeAsO and $LaFeAsO_{0.5}H_{0.5}$. Under chemical doping, the two magnetic parents evolve into SC respectively, which forms a double-dome SC with a minimal $T_c$ at $x\sim0.2$.

In the pressurized $K_xFe_2Se_2$, the $T_c$ is firstly suppressed, and then re-emergent at higher pressure. It is closely related to a so-called quantum critical point. In such two SC, the crystal structures are flexible and the geometry of $FeAs_4$/$FeSe_4$ tetragonal is easy to modify upon chemical doping or pressure to some extent. Thus, the dominant contribution to Fermi surface can shift between Fe-$3d_{xy}$ and Fe-$3d_{xyz}$, and thus the induced Lifshitz transition may in turn determine the SC.

## CONCLUSIONS

We have synthesized a new $Fe_{1.11}Se$ single crystals that exhibit a $T_c$ of 30.4 K utilizing a two-step hydrothermal ion-exchange and de-intercalation process. Our comprehensive structural and spectroscopic analyses confirm the incorporation of 11% interstitial $Fe_2$ ions. The pressure-induced evolution of SC in $Fe_{1.11}Se$ behaves like the FeSe-intercalates rather than binary *s*-FeSe. This work demonstrates the potential of metastable strategy in searching high-$T_c$ superconductors and inducing emergent physical properties in FeSe and other layered superconductors with weak interlayer force.

## METHODS

***Hydrothermal Synthesis of $Fe_{1.11}Se$ Single Crystals:*** $Fe_{1.11}Se$ single crystals were synthesized via a hydrothermal ion-exchange and de-intercalation reaction process involving four sequential steps. Firstly, high-purity tetragonal FeSe powder is obtained by melting Fe and Se pellets together in a quartz tube at 1100 °C, and then annealing at 400 °C for 3-5 days. Then, high-quality single crystals of $K_{0.8}Fe_2Se_2$ were synthesized based on the methods described in Reference.[16] Subsequently, the as-grown $K_{0.8}Fe_2Se_2$ crystals, iron powders, and LiOH pellets were reacted in an aqueous medium within a sealed autoclave under high temperature and pressure, as described in Reference.[55] The golden-yellow $(Li_{1-y}Fe_y)OHFeSe$ ($y$=0.15) single crystals were washed and collected. Finally, the $(Li_{1-y}Fe_y)OHFeSe$ crystals were placed in a Teflon-lined autoclave with 0.1 g KOH, 0.3 g Fe powder, 1 g Sn powder, and 10 mL deionized water. The mixture was heated at 120°C for three days. The final $Fe_{1.11}Se$ crystals were collected from the aqueous solution, and surface water was removed by vacuum evacuation. It yields large, air-stable, silver-white $Fe_{1.11}Se$ single crystals.

***Structural and Chemical Characterization:*** Powder XRD measurements were performed on a Rigaku SmartLab diffractometer to evaluate the phase purity and orientation of single-crystal specimens. Chemical composition analysis was conducted using energy-dispersive X-ray spectroscopy (EDX) on pristine crystal surfaces and inductively coupled plasma atomic emission spectroscopy (ICP-AES) on acid-dissolved samples. For structural determination, single-crystal X-ray diffraction (XRD) measurement was carried out at 299 K using a Bruker D8 diffractometer with Mo-Kα radiation ($\lambda$ = 0.71073 Å). Dynamic gas composition analysis of $Fe_{1.11}Se$ single crystal during isothermal heating under high vacuum via Quadrupole Mass Spectrometry (QMS). The local environment and valence state of iron atoms were investigated by $^{57}Fe$ Mössbauer spectroscopy at 23 K.

*In-situ* high-pressure synchrotron diffractions were carried out at the BL15U1 station of the Shanghai Synchrotron Radiation Facility (SSRF) using a diamond anvil cell (DAC) with a facet diameter of 300 μm. The wavelength of the synchrotron beam is 0.6199 Å. Mineral oil served as the pressure-transmitting medium, and the pressure was calibrated by the standard ruby fluorescence method.

**Magnetic and Electrical Transport measurements:** Magnetic properties were measured using a superconducting quantum interference device (SQUID) magnetometer under an applied field of 10 Oe. Temperature-dependent electrical resistivity was measured using a standard four-probe configuration with applied currents below 2 mA. Hall coefficients ($R_H$) were determined from linear fits of the Hall resistivity ($\rho_{xy}$) versus magnetic field ($H$) in the range of -5 to +5 T at temperatures of 20, 40, 60, 80, 120, 160, 200, and 250 K. *In-situ* high-pressure electrical transports were measured in a physical property measurement system (PPMS, Quantum Design, DynaCool). All measurements were carried out using a beryllium copper alloy DAC with a facet diameter of 500 μm, and pressure was monitored with the ruby fluorescence method. We use KBr as both the pressure-transmitting and insulating medium.

**DFT Calculation:** First-principles calculations were carried out with the density functional theory (DFT) implemented in the Vienna ab initio simulation package (VASP).[56] The generalized gradient approximation (GGA) in the form of Perdew-Burke-Ernzerhof (PBE) was adopted for the exchange-correlation potentials[57]. We used the projector augmented-wave (PAW).[58,59] Because of random occupation of interstitial Fe2, we performed the calculations in a 2×2×1 supercell of FeSe (*P4/nmm*) and introduced 12.5% interstitial Fe to model $Fe_9Se_8$. The planewave energy cut-off was set to 900 eV, and the energy convergence criterion was set to 0.1 eV, using a $\Gamma$-centered 9×9×12 $k$-mesh for Brillouin-zone sampling.

## ASSOCIATED CONTENT

**Supporting Information**

The Supporting Information is available free of charge at ACS Publications website.
Tables S1 and S2 show the refinement results by SCXRD and PXRD, respectively. Figures show $\rho$-$T$ curves under fields, magnetic susceptibility, pressure-dependent $\rho$-$T$ curves for sulfur-doped samples, contour plots and fitting curves for synchrotron diffraction, details for Hall resistivity of all samples, band structure and Fermi surface.

## AUTHOR INFORMATION

**Corresponding Author**

**Jian-gang Guo** - Beijing National Laboratory for Condensed Matter Physics, Institute of Physics, Chinese Academy of Sciences, Beijing 100190, China; orcid.org/0000-0003-3880-3012; Email: jgguo@iphy.ac.cn

**Shifeng Jin** - Beijing National Laboratory for Condensed Matter Physics, Institute of Physics, Chinese Academy of Sciences, Beijing 100190, China; Hefei National Lab., Hefei 230088, China; School of Physical Sciences, University of Chinese Academy of Sciences, Beijing 100049, China; orcid.org/0000-0002-3522-1060; Email: shifengjin@iphy.ac.cn

**Xiaolong Chen** - Beijing National Laboratory for Condensed Matter Physics, Institute of Physics, Chinese Academy of Sciences, Beijing 100190, China; School of Physical Sciences, University of Chinese Academy of Sciences, Beijing 100049, China; orcid.org/0000-0001-8455-2117; Email: xlchen@iphy.ac.cn

**Authors**

**Mingzhang Yang** - Beijing National Laboratory for Condensed Matter Physics, Institute of Physics, Chinese Academy of Sciences, Beijing 100190, China; College of Materials Science and Opto-Electronic Technology, University of Chinese Academy of Sciences, Beijing 101408, China;

**Yuxin Ma** - Beijing National Laboratory for Condensed Matter Physics, Institute of Physics, Chinese Academy of Sciences, Beijing 100190, China; School of Science, China University of Geosciences, Beijing (CUGB), 100083 Beijing, China;

**Qi Li** - Beijing National Laboratory for Condensed Matter Physics, Institute of Physics, Chinese Academy of Sciences, Beijing 100190, China; School of Physical Sciences, University of Chinese Academy of Sciences, Beijing 100049, China;

**Ke Ma** - Beijing National Laboratory for Condensed Matter Physics, Institute of Physics, Chinese Academy of Sciences, Beijing 100190, China; School of Physical Sciences, University of Chinese Academy of Sciences, Beijing 100049, China;

**Jiali Lu** - Beijing National Laboratory for Condensed Matter Physics, Institute of Physics, Chinese Academy of Sciences, Beijing 100190, China; College of Materials Science and Opto-Electronic Technology, University of Chinese Academy of Sciences, Beijing 101408, China;

**Zhaolong Liu** - Beijing National Laboratory for Condensed Matter Physics, Institute of Physics, Chinese Academy of Sciences, Beijing 100190, China; College of Materials Science and Opto-Electronic Technology, University of Chinese Academy of Sciences, Beijing 101408, China;

**Ruijin Sun** - School of Science, China University of Geosciences, Beijing (CUGB), 100083 Beijing, China;

**Tianping Ying** - Beijing National Laboratory for Condensed Matter Physics, Institute of Physics, Chinese Academy of Sciences, Beijing 100190, China;

**Mengdi Wang** - Key Laboratory of Quantum Materials under Extreme Conditions in Shandong Province, School of Physics and Physical Engineering, Qufu Normal University, Qufu 273165, China;

**Xin Chen** - Key Laboratory of Quantum Materials under Extreme Conditions in Shandong Province, School of Physics and Physical Engineering, Qufu Normal University, Qufu 273165, China;

**Changchun Zhao** - School of Science, China University of Geosciences, Beijing (CUGB), 100083 Beijing, China;
Complete contact information is available at:
https://pubs.acs.org/

**Author Contributions**

**Notes**

The authors declare no competing financial interest.

## ACKNOWLEDGMENT

We thank Jiazhen Zhang for discussions. We thank Youting Song and Ziyan Lv for SCXRD measurement. This work is financially supported by the National Key Research and Development Program of China (2023YFA1406301), Quantum Science and Technology-National Science and Technology Major Project 2021ZD0302704, the National Natural Science Foundation of China (52525205, 52272268, 52250308). This work was supported by the Synergetic Extreme Condition User Facility (SECUF), including High-pressure Synergetic Measurement Station and the Sample Pre-selection and Characterization Station. We thank the staff from Shanghai Synchrotron Radiation Facility (SSRF) at BL15U.

# REFERENCES

(1) Hsu, F.-C.; Luo, J.-Y.; Yeh, K.-W.; Chen, T.-K.; Huang, T.-W.; Wu, P. M.; Lee, Y.-C.; Huang, Y.-L.; Chu, Y.-Y.; Yan, D.-C.; Wu, M.-K. Superconductivity in the PbO-Type Structure α-FeSe. *Proc. Natl. Acad. Sci. U.S.A* **2008**, *105* (38), 14262–14264.

(2) Margadonna, S.; Takabayashi, Y.; T. McDonald, M.; Kasperkiewicz, K.; Mizuguchi, Y.; Takano, Y.; N. Fitch, A.; Suard, E.; Prassides, K. Crystal Structure of the New FeSe₁₋ₓ Superconductor. *Chem. Commun.* **2008**, *0* (43), 5607–5609.

(3) Liu, D.; Zhang, W.; Mou, D.; He, J.; Ou, Y.-B.; Wang, Q.-Y.; Li, Z.; Wang, L.; Zhao, L.; He, S.; Peng, Y.; Liu, X.; Chen, C.; Yu, L.; Liu, G.; Dong, X.; Zhang, J.; Chen, C.; Xu, Z.; Hu, J.; Chen, X.; Ma, X.; Xue, Q.; Zhou, X. J. Electronic Origin of High-Temperature Superconductivity in Single-Layer FeSe Superconductor. *Nat. Commun.* **2012**, *3* (1), 931.

(4) Ge, J.-F.; Liu, Z.-L.; Liu, C.; Gao, C.-L.; Qian, D.; Xue, Q.-K.; Liu, Y.; Jia, J.-F. Superconductivity above 100 K in Single-Layer FeSe Films on Doped SrTiO₃. *Nat. Mater.* **2015**, *14* (3), 285–289.

(5) Yin, J.-X.; Wu, Z.; Wang, J.-H.; Ye, Z.-Y.; Gong, J.; Hou, X.-Y.; Shan, L.; Li, A.; Liang, X.-J.; Wu, X.-X.; Li, J.; Ting, C.-S.; Wang, Z.-Q.; Hu, J.-P.; Hor, P.-H.; Ding, H.; Pan, S. H. Observation of a Robust Zero-Energy Bound State in Iron-Based Superconductor Fe(Te,Se). *Nat. Phys.* **2015**, *11* (7), 543–546.

(6) Wang, Z. F.; Zhang, H.; Liu, D.; Liu, C.; Tang, C.; Song, C.; Zhong, Y.; Peng, J.; Li, F.; Nie, C.; Wang, L.; Zhou, X. J.; Ma, X.; Xue, Q. K.; Liu, F. Topological Edge States in a High-Temperature Superconductor FeSe/SrTiO₃(001) Film. *Nat. Mater.* **2016**, *15* (9), 968–973.

(7) Zhang, P.; Yaji, K.; Hashimoto, T.; Ota, Y.; Kondo, T.; Okazaki, K.; Wang, Z.; Wen, J.; Gu, G. D.; Ding, H.; Shin, S. Observation of Topological Superconductivity on the Surface of an Iron-Based Superconductor. *Science* **2018**, *360* (6385), 182–186.

(8) McQueen, T. M.; Huang, Q.; Ksenofontov, V.; Felser, C.; Xu, Q.; Zandbergen, H.; Hor, Y. S.; Allred, J.; Williams, A. J.; Qu, D.; Checkelsky, J.; Ong, N. P.; Cava, R. J. Extreme Sensitivity of Superconductivity to Stoichiometry in Fe₁₊δSe. *Phys. Rev. B* **2009**, *79* (1), 014522.

(9) Fang, M. H.; Pham, H. M.; Qian, B.; Liu, T. J.; Vehstedt, E. K.; Liu, Y.; Spinu, L.; Mao, Z. Q. Superconductivity Close to Magnetic Instability in Fe(Se₁₋ₓTeₓ)₀.₈₂. *Phys. Rev. B* **2008**, *78* (22), 224503.

(10) Yeh, K.-W.; Huang, T.-W.; Huang, Y.; Chen, T.-K.; Hsu, F.-C.; M. Wu, P.; Lee, Y.-C.; Chu, Y.-Y.; Chen, C.-L.; Luo, J.-Y.; Yan, D.-C.; Wu, M.-K. Tellurium Substitution Effect on Superconductivity of the α-Phase Iron Selenide. *Europhys. Lett.* **2008**, *84* (3), 37002.

(11) Gresty, N. C.; Takabayashi, Y.; Ganin, A. Y.; McDonald, M. T.; Claridge, J. B.; Giap, D.; Mizuguchi, Y.; Takano, Y.; Kagayama, T.; Ohishi, Y.; Takata, M.; Rosseinsky, M. J.; Margadonna, S.; Prassides, K. Structural Phase Transitions and Superconductivity in Fe₁₊δSe₀.₅₇Te₀.₄₃ at Ambient and Elevated Pressures. *J. Am. Chem. Soc.* **2009**, *131* (46), 16944–16952.

(12) Mizuguchi, Y.; Tomioka, F.; Tsuda, S.; Yamaguchi, T.; Takano, Y. Substitution Effects on FeSe Superconductor. *J. Phys. Soc. Jpn.* **2009**, *78* (7), 074712–074712.

(13) Katayama, N.; Ji, S.; Louca, D.; Lee, S.; Fujita, M.; J. Sato, T.; Wen, J.; Xu, Z.; Gu, G.; Xu, G.; Lin, Z.; Enoki, M.; Chang, S.; Yamada, K.; M. Tranquada, J. Investigation of the Spin-Glass Regime between the Antiferromagnetic and Superconducting Phases in Fe₁₊ᵧSeₓTe₁₋ₓ. *J. Phys. Soc. Jpn.* **2010**, *79* (11), 113702.

(14) Hosoi, S.; Matsuura, K.; Ishida, K.; Wang, H.; Mizukami, Y.; Watashige, T.; Kasahara, S.; Matsuda, Y.; Shibauchi, T. Nematic Quantum Critical Point without Magnetism in FeSe₁₋ₓSₓ Superconductors. *Proc. Natl. Acad. Sci. U.S.A* **2016**, *113* (29), 8139–8143.

(15) Licciardello, S.; Buhot, J.; Lu, J.; Ayres, J.; Kasahara, S.; Matsuda, Y.; Shibauchi, T.; Hussey, N. E. Electrical Resistivity across a Nematic Quantum Critical Point. *Nature* **2019**, *567* (7747), 213–217.

(16) Guo, J.; Jin, S.; Wang, G.; Wang, S.; Zhu, K.; Zhou, T.; He, M.; Chen, X. Superconductivity in the Iron Selenide KₓFe₂Se₂ (0<x<1.0). *Phys. Rev. B* **2010**, *82* (18), 180520.

(17) Fang, M.-H.; Wang, H.-D.; Dong, C.-H.; Li, Z.-J.; Feng, C.-M.; Chen, J.; Yuan, H. Q. Fe-Based Superconductivity with Tᶜ=31 K Bordering an Antiferromagnetic Insulator in (Tl,K)FeₓSe₂. *Europhys. Lett.* **2011**, *94* (2), 27009.

(18) Wang, A. F.; Ying, J. J.; Yan, Y. J.; Liu, R. H.; Luo, X. G.; Li, Z. Y.; Wang, X. F.; Zhang, M.; Ye, G. J.; Cheng, P.; Xiang, Z. J.; Chen, X. H. Superconductivity at 32 K in Single-Crystalline Rb₍ₓ₎Fe₂₋ᵧSe₂. *Phys. Rev. B* **2011**, *83* (6), 060512.

(19) Li, C.-H.; Shen, B.; Han, F.; Zhu, X.; Wen, H.-H. Transport Properties and Anisotropy of Rb₁₋ₓFe₂₋ᵧSe₂ Single Crystals. *Phys. Rev. B* **2011**, *83* (18), 184521.

(20) Ying, J. J.; Wang, X. F.; Luo, X. G.; Wang, A. F.; Zhang, M.; Yan, Y. J.; Xiang, Z. J.; Liu, R. H.; Cheng, P.; Ye, G. J.; Chen, X. H. Superconductivity and Magnetic Properties of Single Crystals of K₀.₇₅Fe₁.₆₆Se₂ and Cs₀.₈₁Fe₁.₆₁Se₂. *Phys. Rev. B* **2011**, *83* (21), 212502.

(21) Ying, T. P.; Chen, X. L.; Wang, G.; Jin, S. F.; Zhou, T. T.; Lai, X. F.; Zhang, H.; Wang, W. Y. Observation of Superconductivity at 30~46 K in AₓFe₂Se₂(A = Li, Na, Ba, Sr, Ca, Yb and Eu). *Sci Rep* **2012**, *2* (1), 426.

(22) Ying, T.; Chen, X.; Wang, G.; Jin, S.; Lai, X.; Zhou, T.; Zhang, H.; Shen, S.; Wang, W. Superconducting Phases in Potassium-Intercalated Iron Selenides. *J. Am. Chem. Soc.* **2013**, *135* (8), 2951–2954.

(23) Guo, J.; Lei, H.; Hayashi, F.; Hosono, H. Superconductivity and Phase Instability of NH₃-Free Na-Intercalated FeSe₁₋ₓSₓ. *Nat. Commun.* **2014**, *5* (1), 4756.

(24) Hayashi, F.; Lei, H.; Guo, J.; Hosono, H. Modulation Effect of Interlayer Spacing on the Superconductivity of Electron-Doped FeSe-Based Intercalates. *Inorg. Chem.* **2015**, *54* (7), 3346–3351.

(25) Cassidy, S. J.; Woodruff, D. N.; Sedlmaier, S. J.; Blandy, J. N.; Reinhard, C.; Magdysyuk, O. V.; Goodwin, A. L.; Ramos, S.; Clarke, S. J. Stepwise Reactions in the Potassium and Ammonia-Intercalated Iron Selenide Superconductor Phase Diagram Followed by In Situ Powder Diffraction. *J. Am. Chem. Soc.* **2025**, *147* (22), 18563–18575.

(26) Lu, X. F.; Wang, N. Z.; Wu, H.; Wu, Y. P.; Zhao, D.; Zeng, X. Z.; Luo, X. G.; Wu, T.; Bao, W.; Zhang, G. H.; Huang, F. Q.; Huang, Q. Z.; Chen, X. H. Coexistence of Superconductivity and Antiferromagnetism in (Li₀.₈Fe₀.₂)OHFeSe. *Nat. Mater.* **2015**, *14* (3), 325–329.

(27) Dong, X.; Zhou, H.; Yang, H.; Yuan, J.; Jin, K.; Zhou, F.; Yuan, D.; Wei, L.; Li, J.; Wang, X.; Zhang, G.; Zhao, Z. Phase Diagram of (Li₁₋ₓFeₓ)OHFeSe: A Bridge between Iron Selenide and Arsenide Superconductors. *J. Am. Chem. Soc.* **2015**, *137* (1), 66–69.

(28) Wang, Q.-Y.; Li, Z.; Zhang, W.-H.; Zhang, Z.-C.; Zhang, J.-S.; Li, W.; Ding, H.; Ou, Y.-B.; Deng, P.; Chang, K.; Wen, J.; Song,

C.-L.; He, K.; Jia, J.-F.; Ji, S.-H.; Wang, Y.-Y.; Wang, L.-L.; Chen, X.; Ma, X.-C.; Xue, Q.-K. Interface-Induced High-Temperature Superconductivity in Single Unit-Cell FeSe Films on $SrTiO_3$. *Chin. Phys. Lett.* **2012**, *29* (3), 037402.

(29) Lee, J. J.; Schmitt, F. T.; Moore, R. G.; Johnston, S.; Cui, Y.-T.; Li, W.; Yi, M.; Liu, Z. K.; Hashimoto, M.; Zhang, Y.; Lu, D. H.; Devereaux, T. P.; Lee, D.-H.; Shen, Z.-X. Interfacial Mode Coupling as the Origin of the Enhancement of $T_c$ in FeSe Films on $SrTiO_3$. *Nature* **2014**, *515* (7526), 245–248.

(30) Mizuguchi, Y.; Tomioka, F.; Tsuda, S.; Yamaguchi, T.; Takano, Y. Superconductivity at 27K in Tetragonal FeSe under High Pressure. *Appl. Phys. Lett.* **2008**, *93* (15), 152505.

(31) Medvedev, S.; McQueen, T. M.; Troyan, I. A.; Palasyuk, T.; Eremets, M. I.; Cava, R. J.; Naghavi, S.; Casper, F.; Ksenofontov, V.; Wortmann, G.; Felser, C. Electronic and Magnetic Phase Diagram of $\beta$-$Fe_{1.01}$Se with Superconductivity at 36.7 K under Pressure. *Nat. Mater.* **2009**, *8* (8), 630–633.

(32) Sun, J. P.; Matsuura, K.; Ye, G. Z.; Mizukami, Y.; Shimozawa, M.; Matsubayashi, K.; Yamashita, M.; Watashige, T.; Kasahara, S.; Matsuda, Y.; Yan, J.-Q.; Sales, B. C.; Uwatoko, Y.; Cheng, J.-G.; Shibauchi, T. Dome-Shaped Magnetic Order Competing with High-Temperature Superconductivity at High Pressures in FeSe. *Nat. Commun.* **2016**, *7* (1), 12146.

(33) Sun, J. P.; Ye, G. Z.; Shahi, P.; Yan, J.-Q.; Matsuura, K.; Kontani, H.; Zhang, G. M.; Zhou, Q.; Sales, B. C.; Shibauchi, T.; Uwatoko, Y.; Singh, D. J.; Cheng, J.-G. High-$T_c$ Superconductivity in FeSe at High Pressure: Dominant Hole Carriers and Enhanced Spin Fluctuations. *Phys. Rev. Lett.* **2017**, *118* (14), 147004.

(34) Wang, P. S.; Sun, S. S.; Cui, Y.; Song, W. H.; Li, T. R.; Yu, R.; Lei, H.; Yu, W. Pressure Induced Stripe-Order Antiferromagnetism and First-Order Phase Transition in Fese. *Phys. Rev. Lett.* **2016**, *117* (23), 237001.

(35) Kothapalli, K.; Böhmer, A. E.; Jayasekara, W. T.; Ueland, B. G.; Das, P.; Sapkota, A.; Taufour, V.; Xiao, Y.; Alp, E.; Bud'ko, S. L.; Canfield, P. C.; Kreyssig, A.; Goldman, A. I. Strong Cooperative Coupling of Pressure-Induced Magnetic Order and Nematicity in FeSe. *Nat. Commun.* **2016**, *7* (1), 12728.

(36) Sun, L.; Chen, X.-J.; Guo, J.; Gao, P.; Huang, Q.-Z.; Wang, H.; Fang, M.; Chen, X.; Chen, G.; Wu, Q.; Zhang, C.; Gu, D.; Dong, X.; Wang, L.; Yang, K.; Li, A.; Dai, X.; Mao, H.; Zhao, Z. Re-Emerging Superconductivity at 48 Kelvin in Iron Chalcogenides. *Nature* **2012**, *483* (7387), 67–69.

(37) Guo, J.; Chen, X.-J.; Dai, J.; Zhang, C.; Guo, J.; Chen, X.; Wu, Q.; Gu, D.; Gao, P.; Yang, L.; Yang, K.; Dai, X.; Mao, H.; Sun, L.; Zhao, Z. Pressure-Driven Quantum Criticality in Iron-Selenide Superconductors. *Phys. Rev. Lett.* **2012**, *108* (19), 197001.

(38) Sun, J. P.; Shahi, P.; Zhou, H. X.; Huang, Y. L.; Chen, K. Y.; Wang, B. S.; Ni, S. L.; Li, N. N.; Zhang, K.; Yang, W. G.; Uwatoko, Y.; Xing, G.; Sun, J.; Singh, D. J.; Jin, K.; Zhou, F.; Zhang, G. M.; Dong, X. L.; Zhao, Z. X.; Cheng, J.-G. Reemergence of High-$T_c$ Superconductivity in the $(Li_{1-x}Fe_x)$OHFe$_{1-y}$Se under High Pressure. *Nat. Commun.* **2018**, *9* (1), 380.

(39) Shahi, P.; Sun, J. P.; Wang, S. H.; Jiao, Y. Y.; Chen, K. Y.; Sun, S. S.; Lei, H. C.; Uwatoko, Y.; Wang, B. S.; Cheng, J.-G. High-$T_c$ Superconductivity up to 55 K under High Pressure in a Heavily Electron Doped $Li_{0.36}(NH_3)_yFe_2Se_2$ Single Crystal. *Phys. Rev. B* **2018**, *97* (2), 020508.

(40) Wang, X. C.; Liu, Q. Q.; Lv, Y. X.; Gao, W. B.; Yang, L. X.; Yu, R. C.; Li, F. Y.; Jin, C. Q. The Superconductivity at 18 K in LiFeAs System. *Solid State Commun.* **2008**, *148* (11), 538–540.

(41) Kiiamov, A. G.; Lysogorskiy, Y. V.; Vagizov, F. G.; Tagirov, L. R.; Tayurskii, D. A.; Croitori, D.; Tsurkan, V.; Loidl, A. Mössbauer Spectroscopy Evidence of Intrinsic Non-Stoichiometry in Iron Telluride Single Crystals. *Ann. Phys.* **2017**, *529* (4), 1600241.

(42) Kiiamov, A. G.; Tayurskii, D. A.; Vagizov, F. G.; Croitori, D.; Tsurkan, V.; Krug von Nidda, H.-A.; Tagirov, L. R. DFT and Mössbauer Spectroscopy Study of a $FeTe_{0.5}Se_{0.5}$ Single Crystal. *JETP Lett.* **2019**, *109* (4), 266–269.

(43) Sun, S.; Wang, S.; Yu, R.; Lei, H. Extreme Anisotropy and Anomalous Transport Properties of Heavily Electron Doped $Li_x(NH_3)_yFe_2Se_2$ Single Crystals. *Phys. Rev. B* **2017**, *96* (6), 064512.

(44) Wang, Z.; Yuan, J.; Wosnitza, J.; Zhou, H.; Huang, Y.; Jin, K.; Zhou, F.; Dong, X.; Zhao, Z. The Upper Critical Field and Its Anisotropy in $(Li_{1-x}Fe_x)$OHFe$_{1-y}$Se. *J. Phys.: Condens. Matter* **2017**, *29* (2), 025701.

(45) Vedeneev, S. I.; Piot, B. A.; Maude, D. K.; Sadakov, A. V. Temperature Dependence of the Upper Critical Field of FeSe Single Crystals. *Phys. Rev. B* **2013**, *87* (13), 134512.

(46) Zhou, N.; Sun, Y.; Xi, C. Y.; Wang, Z. S.; Zhang, J. L.; Zhang, Y.; Zhang, Y. F.; Xu, C. Q.; Pan, Y. Q.; Feng, J. J.; Meng, Y.; Yi, X. L.; Pi, L.; Tamegai, T.; Xing, X.; Shi, Z. Disorder-Robust High-Field Superconducting Phase of FeSe Single Crystals. *Phys. Rev. B* **2021**, *104* (14), L140504.

(47) Ma, X.; Li, K.; Hou, S.; Mei, H.; Dong, Y.; Tan, M.; Xie, M.; Ren, W.; Jiang, X.; Li, Z.; Zhang, A.; Zhang, Q. Distinct Electronic Origins of Superconductivity and Nematicity in Fese. *Phys. Rev. Lett.* **2025**, *134* (25), 256002.

(48) Torchetti, D. A.; Fu, M.; Christensen, D. C.; Nelson, K. J.; Imai, T.; Lei, H. C.; Petrovic, C. $^{77}$Se NMR Investigation of the $K_xFe_{2-y}Se_2$ Superconductor ($T_c$=33 K). *Phys. Rev. B* **2011**, *83* (10), 104508.

(49) Li, W.; Ding, H.; Deng, P.; Chang, K.; Song, C.; He, K.; Wang, L.; Ma, X.; Hu, J.-P.; Chen, X.; Xue, Q.-K. Phase Separation and Magnetic Order in K-Doped Iron Selenide Superconductor. *Nat. Phys.* **2012**, *8* (2), 126–130.

(50) Matsuura, K.; Mizukami, Y.; Arai, Y.; Sugimura, Y.; Maejima, N.; Machida, A.; Watanuki, T.; Fukuda, T.; Yajima, T.; Hiroi, Z.; Yip, K. Y.; Chan, Y. C.; Niu, Q.; Hosoi, S.; Ishida, K.; Mukasa, K.; Kasahara, S.; Cheng, J.-G.; Goh, S. K.; Matsuda, Y.; Uwatoko, Y.; Shibauchi, T. Maximizing $T_c$ by Tuning Nematicity and Magnetism in FeSe$_{1-x}$S$_x$ Superconductors. *Nat. Commun.* **2017**, *8* (1), 1143.

(51) Mukasa, K.; Matsuura, K.; Qiu, M.; Saito, M.; Sugimura, Y.; Ishida, K.; Otani, M.; Onishi, Y.; Mizukami, Y.; Hashimoto, K.; Gouchi, J.; Kumai, R.; Uwatoko, Y.; Shibauchi, T. High-Pressure Phase Diagrams of FeSe$_{1-x}$Te$_x$: Correlation between Suppressed Nematicity and Enhanced Superconductivity. *Nat. Commun.* **2021**, *12* (1), 381.

(52) Izumi, M.; Zheng, L.; Sakai, Y.; Goto, H.; Sakata, M.; Nakamoto, Y.; Nguyen, H. L. T.; Kagayama, T.; Shimizu, K.; Araki, S.; Kobayashi, T. C.; Kambe, T.; Gu, D.; Guo, J.; Liu, J.; Li, Y.; Sun, L.; Prassides, K.; Kubozono, Y. Emergence of Double-Dome Superconductivity in Ammoniated Metal-Doped FeSe. *Sci. Rep.* **2015**, *5* (1), 9477.

(53) Iimura, S.; Matsuishi, S.; Sato, H.; Hanna, T.; Muraba, Y.; Kim, S. W.; Kim, J. E.; Takata, M.; Hosono, H. Two-Dome Structure in Electron-Doped Iron Arsenide Superconductors. *Nat. Commun.* **2012**, *3* (1), 943.

(54) Hiraishi, M.; Iimura, S.; Kojima, K. M.; Yamaura, J.; Hiraka, H.; Ikeda, K.; Miao, P.; Ishikawa, Y.; Torii, S.; Miyazaki, M.; Yamauchi, I.; Koda, A.; Ishii, K.; Yoshida, M.; Mizuki, J.; Kadono, R.; Kumai, R.; Kamiyama, T.; Otomo, T.; Murakami, Y.; Matsuishi, S.; Hosono, H. Bipartite Magnetic Parent Phases in the Iron Oxypnictide Superconductor. *Nature Phys* **2014**, *10* (4), 300–303.

(55) Sun, H.; Woodruff, D. N.; Cassidy, S. J.; Allcroft, G. M.; Sedlmaier, S. J.; Thompson, A. L.; Bingham, P. A.; Forder, S. D.; Cartenet, S.; Mary, N.; Ramos, S.; Foronda, F. R.; Williams, B. H.; Li, X.; Blundell, S. J.; Clarke, S. J. Soft Chemical Control of Superconductivity in Lithium Iron Selenide Hydroxides $Li_{1-x}Fe_x(OH)Fe_{1-y}$Se. *Inorg. Chem.* **2015**, *54* (4), 1958–1964.

(56) Kresse, G.; Furthmüller, J. Efficient Iterative Schemes for Ab Initio Total-Energy Calculations Using a Plane-Wave Basis Set. *Phys. Rev. B* **1996**, *54* (16), 11169–11186.

(57)   Perdew, J. P.; Burke, K.; Ernzerhof, M. Generalized Gradient Approximation Made Simple. *Phys. Rev. Lett.* **1996**, 77 (18), 3865–3868.

(58)   Kresse, G.; Joubert, D. From Ultrasoft Pseudopotentials to the Projector Augmented-Wave Method. *Phys. Rev. B* **1999**, 59 (3), 1758–1775.

(59)   Kresse, G.; Furthmüller, J. Efficiency of Ab-Initio Total Energy Calculations for Metals and Semiconductors Using a Plane-Wave Basis Set. *Comput. Mater. Sci.* **1996**, 6 (1), 15–50.

TOC

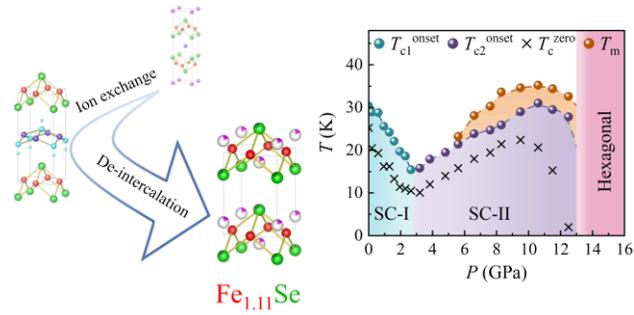